\documentclass[aps, showpacs, prd, superscriptadaddress, twocolumn] {revtex4-1}
\usepackage{hyperref}
 \usepackage{amsmath}
\usepackage{amssymb}
\usepackage{epsfig}
\usepackage{natbib}
\usepackage{epstopdf}
\usepackage{graphicx}
\usepackage[normalem]{ulem}

\usepackage[caption=false,font=Large]{subfig}

\newcommand*{\be}{\begin{equation}}
\newcommand*{\ee}{\end{equation}}
\newcommand*{\bea}{\begin{eqnarray}}
\newcommand*{\eea}{\end{eqnarray}}

\newcommand{\sech}{\, \mathrm{sech}}

 \DeclareFontFamily{OT1}{pzc}{}
 \DeclareFontShape{OT1}{pzc}{m}{it}%
 {<->  s  *  [1.400]  pzcmi7t}{}
\DeclareMathAlphabet{\mathscr}{OT1}{pzc}%
{m}{it}

\begin{document}

\title{Variational formalism   for the Klein-Gordon oscillon}


\author{I. V. Barashenkov and     N. V.  Alexeeva}
 \affiliation{
 Centre for Theoretical  and Mathematical Physics,  University of Cape Town, Rondebosch 7701, South Africa 
 }

\begin{abstract}
The variational method employing the amplitude and width as collective coordinates of the  Klein-Gordon oscillon leads to a
dynamical system with unstable periodic orbits that  blow up when perturbed.
We propose a multiscale variational approach free from the blow-up singularities. An essential feature of 
the  proposed trial function  is the inclusion of the third collective variable: a correction for the nonuniform phase growth. 
In addition to determining the parameters of the oscillon, our approach detects the onset of its instability.
\end{abstract}

\maketitle

\section{Introduction}

 Oscillon is a classical solution describing a
 long-lived localised pulsating
 structure of finite amplitude.  Oscillons 
 play a role in
 the dynamics of inflationary reheating,  symmetry-breaking phase transitions, and  false vacuum decay
  \cite{CGM,Riotto,GInt,11cosmo,Dymnikova,Broadhead,bubbling,Amin1,Stamatopoulos,Zhou,Amin2,Adshead,GG,Bond,Antusch,Hong,Cyn,LozAm}.
  They
occur in the Einstein-Klein-Gordon equations \cite{Maslov,Zhang2,Nazari,Kou1,Hira,Kou2},
 axion models \cite{Kolb,Vaquero,Kawa_axion,Olle,Arvanitaki,Miyazaki},  string phenomenology \cite{string,Kasu,Sang}  and 
   bosonic sector of the standard model 
 \cite{Farhi2,  Graham, Gleiser4, Sfakianakis}. 
 The (2+1)-dimensional oscillons have been studied
  in  the context of the planar Abelian Higgs theory \cite{GT2,Achi}.

 Oscillons were discovered \cite{Voronov1,BM1,BM2,G1}
 in  the (3+1)-dimensional  $\Phi^4$ model,
 \be
\Phi_{tt} -\Delta \Phi - \Phi + \Phi^3=0.
\label{A0}
\ee
The model, together with its (1+1)-dimensional counterpart, remains 
  a workhorse of  quantum field theory \cite{QFT_books_1,QFT_books_2, QFT_books_3, Rychkov, Bajnok, Serone, Graham_Weigel, Martin, Ito, QFT_papers_3D} and cosmology 
\cite{cosmology}.
Despite the apparent simplicity of  equation \eqref{A0}, many properties of its oscillon solution have still not been fully understood
 \cite{Fodor3}.

 Most of the  mathematical analysis of  oscillons has been carried out using asymptotic \cite{Fodor2,Fodor3,Fodor4} and
 numerical  techniques \cite{BM2,G1,CGM,Honda,Gleiser10,Fodor1,Alex_PRD,Fodor3}
 while qualitative insights called on  variational arguments.
 In Ref \cite{CGM},  the $\Phi^4$ oscillon  was  approximated 
 by a localised waveform 
\be
\Phi= 1+A e^{-(r/b)^2},
\label{B1}  \ee
where $A(t)$ is an unknown oscillating amplitude and $b$ is an arbitrarily chosen value of the width.  (Ref \cite{Kev} followed a 
similar strategy when dealing with the two-dimensional sine-Gordon equation.) Once  the ansatz \eqref{B1} has been substituted  in the lagrangian and  the $r$-dependence 
integrated away, the variation
 of action produces 
a second-order equation for  $A(t)$. 

The variational  method does not suggest any optimisation strategies for 
  $b$. 
Making $b(t)$ another collective coordinate  --- as it is done in the studies of the nonlinear Schr\"odinger solitons \cite{Malomed,BAZ} ---
gives rise to an ill-posed dynamical system  not amenable to numerical simulations.
 (See section \ref{naive} below.)

With an obstacle encountered in (3+1) dimensions, one turns to a (1+1) dimensional version of the model for  guidance.
The analysis can be  further simplified by considering oscillons 
approaching
a symmetric vacuum as $x \to \pm \infty$. A physically relevant model of this kind was  considered by Kosevich and Kovalev \cite{KK}:
\be
\phi_{tt} - \phi_{xx} + 4 \phi - 2 \phi^3=0.
\label{A1}
\ee
 Unlike its $\Phi^4$ counterpart, 
the oscillon in the Kosevich-Kovalev  model satisfies $\phi \to 0$ as $x \to \pm \infty$
and oscillates, symmetrically,  between positive and negative values.  
The asymptotic representation of this solution is
\begin{align} 
\phi= \frac{2 \epsilon}{\sqrt 3}   \, \cos (\omega  t )  \, \mathrm{sech} (\epsilon x)
 -\frac{\epsilon^3}{24 \sqrt 3}   \,   \cos ( 3\omega  t) 
  \nonumber \\ \times \mathrm{sech^3} (\epsilon x) 
 +O(\epsilon^5),
\label{A2}
\end{align}
where  $\omega^2 = 4-\epsilon^2$ and $\epsilon \to 0$ \cite{KK}. 
Despite the difference in  the vacuum symmetry, equations \eqref{A0} and \eqref{A1} belong to the same,  Klein-Gordon, variety and  share a number of analytical properties.

The purpose of the present study is to identify a  set of collective coordinates 
and formulate a variational description of the Klein-Gordon oscillon. 
A consistent variational formulation would determine the stability range of the oscillon, 
uncover its instability mechanism and explain some of its properties such as 
the amplitude-frequency relationship. Using  the (1+1)-dimensional Kosevich-Kovalev equation \eqref{A1} as a prototype system, 
we transplant the idea of multiple time scales to the collective-coordinate 
Lagrangian method.  With some modifications,   our approach should remain applicable to oscillons in the (3+1)-dimensional $\Phi^4$ theory 
and other Klein-Gordon models.

Before outlining the paper,  three remarks are in order.

First, 
equation \eqref{A1}  can be seen as  a truncation  of the sine-Gordon model.
The fundamental difference between the Kosevich-Kovalev oscillon and the
sine-Gordon  breather is that the latter solution is exactly periodic 
while the amplitude of the former one decreases due to the third-harmonic 
 radiation. (When the amplitude of the oscillations is small, the radiation is exponentially weak though; hence the decay is slow.)

Second,  it is appropriate to mention an
 alternative variational procedure \cite{Wattis1} where one not only chooses  the spatial part  but also imposes the time dependence of the trial function. For instance, one may set
\[
\phi= A_0 \cos (\omega t) e^{-(r/b)^2}.
\]
For a fixed $\omega$, 
the action  becomes a function of two time-independent parameters, $A_0$ and $b$. 
The shortcoming of this technique is that 
 it does not allow one to examine the stability of the Klein-Gordon oscillon.
 Neither would it capture a slow modulation of the oscillation frequency ---such 
 as the one  
  observed in numerical simulations of the $\Phi^4$ model \cite{BM2,Honda,Fodor1}. 
  
Our last remark concerns a 
 closely related system, the nonlinear Schr\"odinger equation. 
 The variational method has been highly successful in the studies of the Schr\"odinger solitons --- scalar and vector ones,
 with a variety of nonlinearities,  perturbations, and in various dimensions \cite{Malomed}. Several sets of 
 collective coordinates for the Schr\"odinger solitons have been identified. 
 It is the remarkable simplicity and versatility of the variational method demonstrated in the nonlinear Schr\"odinger domain 
  that motivate our search for its Klein-Gordon counterpart.
  
 The outline of the paper is as follows.
  In the next section we show that choosing the collective coordinates similar to the way they are 
  chosen for the nonlinear Schr\"odinger soliton leads to singular finite-dimensional dynamics.
  A consistent variational procedure involving fast and slow temporal scales is formulated in section \ref{multiscale}.
  We assess the approximation  by comparing the 
  variational solution   to the ``true" oscillon obtained numerically.
  Section \ref{Two_remarks} adds remarks on the role of the third collective coordinate
  and the choice of the trial function,
   while an explicit construction of the oscillon with adiabatically 
   changing parameters has been relegated to the Appendix \ref{App}. 
  Finally,  section \ref{conclusions}  summarises conclusions of this study.


\section{Singular amplitude-width dynamics} 
\label{naive}

\subsection{Two-mode variational approximation}

The variational approach to  equation \eqref{A1} makes use of
its Lagrangian,
\be
\label{A3}
L= \frac12 \int \left(\phi_t^2-\phi_x^2-4 \phi^2 +\phi^4  \right) dx.
\ee
Modelling on the nonlinear Schr\"odinger construction \cite{Malomed,BAZ}, we 
choose the amplitude and width  of the oscillon as two collective variables:
\be 
\phi= A \,  \mathrm{sech}  \left(   \frac{x}{b}   \right).
\label{A4}
\ee
The amplitude $A(t) $ is expected to oscillate between positive and negative values
while the width (``breadth") 
$b(t)$ should remain positive at all times. 
Substituting   the Ansatz   \eqref{A4} in \eqref{A3} gives the Lagrangian of a system with two degrees of freedom:
\begin{align} 
L= {\dot A}^2 b +\left( \frac13+ \frac{\pi^2}{36} \right) \frac{ {\dot b}^2 A^2}{b} + \dot A \dot b A- 
\frac{A^2}{3b}   \nonumber   \\ + b \left( \frac23 A^4- 4A^2 \right).
\label{A5}
\end{align} 
In \eqref{A5}, the overdot stands for the derivative with respect to  $t$. 
The equations of motion are  
\begin{subequations} \label{unsys}
\begin{align}
\ddot A+ 4A - \left( \frac13+\frac{\pi^2}{36} \right)  \frac{{\dot b}^2}{b^2} A=
\left(2 \sigma+\frac43 \right)  A^3    \nonumber \\  
 - \left( 2 \sigma +\frac13 \right) \frac{A}{b^2}
\label{A6b} 
\end{align}
and 
\be
\ddot b + 2 \frac{\dot A}{A} \dot b = 4\sigma  \left( \frac{1}{b^2}- A^2 \right) b, \label{A6a} 
\ee
\end{subequations}
where we have introduced a short-hand notation for a numerical factor
\[
\sigma = \frac{1}{1+ \pi^2/3}.    
\]

\subsection{Asymptotic solution}

The system  \eqref{unsys} has a family of  periodic solutions. 
For reasons that will become clear in what follows, 
these solutions are difficult to obtain by means of numerical simulations of equations \eqref{unsys}.
However the family can be constructed as a multiscale perturbation expansion --- in the limit of 
  small $A$ and  large $b$. 
  
  To this end, we let
\be 
A= \epsilon A_1+ \epsilon^3 A_3 +..., \quad 
b= \frac{1}{\epsilon} + \epsilon b_1 + ... , 
\label{A7}
\ee
where $A_1, A_3, ...$ and $b_1, b_3, ....$ are functions of a sequence of temporal variables ${\mathcal T_0} , {\mathcal T_2} , ...$, with
${\mathcal T_{2n}} = \epsilon^{2n } t$ and $\epsilon \to 0$.  
Writing $d/dt= \partial / \partial {\mathcal T_0} + \epsilon^2 \partial {\mathcal T_2}  + ...$ and
substituting the expansions \eqref{A7} in \eqref{A6b},
we set coefficients of like powers of $\epsilon$ to zero.   

The order $\epsilon^1$ gives a linear equation
\[
\frac{\partial^2 A_1}{\partial {\mathcal T_0} ^2}  + 4A_1=0.
\]
Without loss of generality we can take a solution in the form 
\be
A_1=  \psi e^{  2i{\mathcal T_0} } + c.c. = 2 |\psi| \cos  2({\mathcal T_0} -  \theta),
\label{A9}
\ee
where $\psi= \psi({\mathcal T_2} , ....)=|\psi| e^{ -2i \theta}$ is a complex-valued function of ``slow" variables. 
The next order, $\epsilon^3$, gives 
\begin{align} 
\frac{\partial^2 A_3}{ \partial {\mathcal T_0} ^2} +4A_3= - 2 \frac{\partial^2 A_1}{\partial {\mathcal T_0}  \partial {\mathcal T_2} } -  \left(2 \sigma+\frac13 \right) A_1  \nonumber  \\ 
+ \left(2 \sigma + \frac43 \right) A_1^3.
\label{A10}
\end{align} 
Substituting for $A_1$ from \eqref{A9} and imposing the nonsecularity condition 
\be
4i \frac{\partial \psi}{\partial {\mathcal T_2} } +
\left( 2\sigma+\frac13 \right) \psi - (6 \sigma +4)  \psi |\psi|^2=0,
\label{A11}
\ee
we determine
 a  solution of \eqref{A10}:
\be
A_3=-\frac{1}{8}  \left( \sigma+\frac23 \right)    | \psi|^3  \cos 6({\mathcal T_0} -\theta).
\label{A12}
\ee

          \begin{figure}[t]
 \begin{center} 
        \includegraphics*[width=\linewidth] {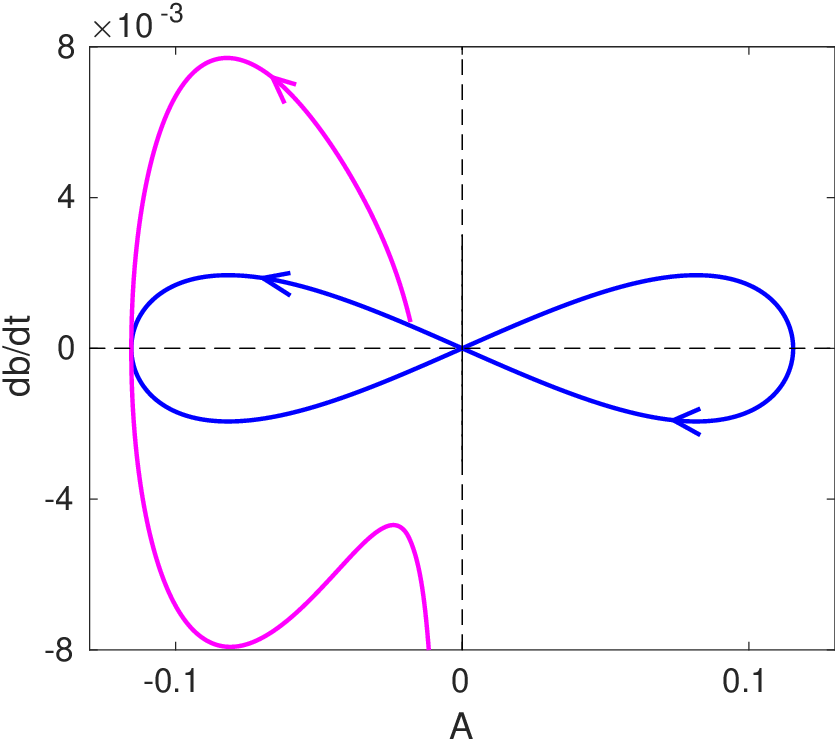}   
                     \end{center}
  \caption{Trajectories  of the  four-dimensional system \eqref{unsys} projected on the $(A, \dot b)$ plane. The $\infty$-shaped curve 
  describes the periodic solution \eqref{21} with $\epsilon=0.1$. The magenta curve depicts a solution evolving from the initial conditions taken
  slightly off the periodic trajectory. 
  The initial values $A(0)$, $b(0)$ and $\dot A(0)$
  for this perturbation are given by the first two terms in equations \eqref{A18} and \eqref{A19}, 
  with $\epsilon=0.1$ and  $t=t_0= 0.55 \pi/ \omega$. 
  The initial condition for $\dot b$ is  $\dot b(0)= -\frac{\sigma}{24}  \epsilon  \omega \sin(2 \omega t_0)+10^{-4}$, 
  with the same $\epsilon, \omega$ and $t_0$. 
     \label{PhasePortrait}
  }
 \end{figure}

Turning to equation \eqref{A6a}, the leading order is
\be
\frac{\partial^2 b_1}{\partial {\mathcal T_0} ^2} + \frac{2 }{ A_1}            \frac{\partial  A_1}{\partial {\mathcal T_0} } \frac {\partial b_1}{\partial {\mathcal T_0} } = \sigma (1-A_1^2).
\label{A13}
\ee
The general solution of this linear equation is given by
\begin{align}
b_1=  \frac{\sigma}{4} (1- 3 |\psi|^2) \tau  \tan 2 \tau 
+\frac{\sigma}{16} |\psi|^2 \cos 4 \tau   \nonumber \\
+ \frac{C_1}{2}  \tan 2 \tau,
\label{A21}
\end{align}
where $\tau= \mathcal T_0-\theta$ and 
$C_1$ is an arbitrary constant in front of a homogeneous solution.
(The second homogeneous solution was absorbed in the term $1/\epsilon$
 in the expansion \eqref{A7}.)
Letting $C_{1}=0$ and imposing the constraint
\be
1- 3 |\psi|^2=0
\label{A14}
\ee
selects a regular solution:
\be
b_1=\frac{\sigma }{48}  \cos 4 \tau.
\label{A15}
\ee

Finally, the phase of the  complex variable $\psi$ is determined by equation \eqref{A11}.
Substituting $|\psi|$ from \eqref{A14} we obtain
\[
\theta= \frac18 {\mathcal T_2} .
\]

Thus, the asymptotic solution of equations \eqref{unsys} has the form 
\begin{subequations} \label{21}
\begin{align}
A= \frac{2}{\sqrt 3}   \epsilon \cos \omega t  - \frac{3\sigma + 2}{72 \sqrt 3}     \epsilon^3
 \cos 3 \omega t+O(\epsilon^5),     \label{A18} \\
 b=\frac{1}{\epsilon} + \frac{   \sigma }{48}      \epsilon  \cos 2 \omega t +O(\epsilon^3),  \label{A19}
  \end{align}
where $\epsilon \to 0$ and 
\be 
\omega= 2- \epsilon^2/4+ O(\epsilon^4).
\ee  \end{subequations}
   This solution  describes a closed  orbit  in the phase space of the system \eqref{unsys}.  See Fig \ref{PhasePortrait}.

     \subsection{Singular dynamics}

   It is not difficult to realise that the asymptotic   solution \eqref{21} 
    is unstable.
   Indeed, the bounded solution \eqref{A15} of equation \eqref{A13} is selected by the
   initial condition   $\partial b_1/ \partial {\mathcal T_0} =0$ at ${\mathcal T_0} =\theta$.
   If we, instead,  let $\partial b_1/ \partial {\mathcal T_0} =\delta$ with a small $\delta$, 
   the $\tan 2 \tau$ component will be turned on in the expression \eqref{A21} 
   and $b_1$ will blow up at  ${\mathcal T_0} = \theta+ \pi/4$. 
   Fig \ref{PhasePortrait} illustrates the evolution of a small perturbation of the periodic orbit.

  The numerical analysis of the system \eqref{unsys} 
   indicates that  periodic solutions with $A(t)$
  oscillating about zero are unstable for any value of the oscillation amplitude --- and not only in the small-$A$ asymptotic regime.
  The instability originates from the topology of 
the four-dimensional phase space of the system that features a singularity at $A=0$.

Indeed, had the system not had a  singularity and had the periodic orbit been stable, 
a small perturbation about it would have been oscillating, quasiperiodically, between positive and negative $A$.
The corresponding trajectory would be winding on a torus in the four-dimensional phase space, 
with the points where the trajectory passes through 
$A=0$  filling a  finite interval on the $\dot b$-axis.
In the presence of the singularity, however,  such a torus cannot form because
any trajectory crossing through $A=0$ at time $t_*$  has to satisfy $\dot b=0$ at the same time.  

\begin{figure}[t]
 \begin{center} 
                      \includegraphics*[width=\linewidth] {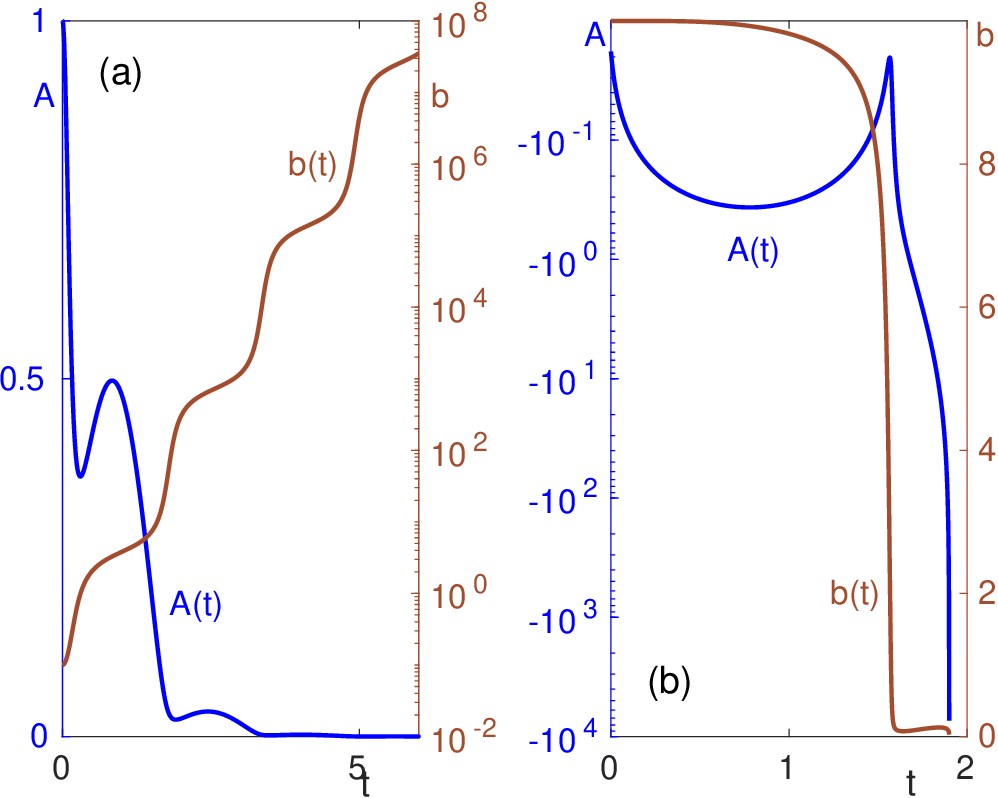}   
                                         \end{center}
  \caption{Two types of unstable evolution in equations \eqref{unsys}. 
  (a) $A(t)$ approaches zero while $b(t)$ grows exponentially.  
  (b)  $A(t)$  grows to infinity (negative infinity in this simulation) while $b(t)$ shrinks to zero.
  \label{double}
  }
 \end{figure}

Trajectories that do not pass through the plane $A=\dot b=0$ follow one of two scenarios.
In the ``spreading" scenario, the width $b(t)$ escapes to infinity (Fig \ref{double}(a)).
The corresponding $A(t)$ approaches zero but remains on one side of it at all times. 
In the alternative scenario,  the amplitude $A(t)$ blows up while the width shrinks to zero  (Fig \ref{double}(b)).

Due to the singularity of solutions emerging from generic initial conditions, the system  \eqref{unsys} is not amenable to  numerical simulations beyond a few oscillation cycles.
What is even more important, the all-$\omega$ universal
 instability of periodic solutions of this four-dimensional system does not match up with the behaviour of the oscillon solutions of the 
  full partial differential equation \eqref{A1}. Contrary to the predictions of the two-mode approximation, the simulations of equation \eqref{A1} demonstrate that the nearly-periodic 
  oscillons with frequencies in the range $\sqrt{2} \lesssim \omega <2$ are stable. 
  The amplitude and frequency of such oscillons do change due to the 
 third-harmonic radiation; however, these changes are 
 slow and may only be noticeable over long temporal intervals.  (See Fig \ref{3D}(a)).

  We note that 
  an  ill-posed system  similar to \eqref{unsys} was encountered in the variational studies of the sine-Gordon breathers \cite{CF}.

The spurious instability of periodic trajectories of the system \eqref{unsys}
 disqualifies the two-variable Ansatz \eqref{A4} 
and prompts one to look for suitable alternatives.

      \section{Multiscale variational method}
   \label{multiscale} 
   \subsection{Amplitude, width and phase correction}

  To rectify the flaws of the ``naive" variational algorithm, we consider 
  $\phi$  to be a function of two  time variables, ${\mathcal T_0}=t$ and 
  ${\mathcal T_1}= \epsilon t$.
    The rate of change  is assumed to be $O(1)$ on either scale:
   $\partial \phi/ \partial \mathcal T_0, \partial \phi/ \partial \mathcal T_1 \sim 1$. 
   We require  $\phi$ to be periodic in $\mathcal T_0$, with a period of $T$:
   \[\mathcal 
   \phi(\mathcal T_0+ T; \,   \mathcal T_1)=\phi(\mathcal T_0; \,  \mathcal T_1).
   \]
   
As $\epsilon \to 0$, the variables $\mathcal T_0 $ and $\mathcal T_1$  become independent and 
   the Lagrangian \eqref{A3} transforms to 
   \begin{align} 
L=  \int \left[   \left( \frac{\partial  \phi}{\partial {\mathcal T_0} } + \epsilon \frac{\partial  \phi}{\partial {\mathcal T_1} }   \right)^2
-\phi_x^2   
-4 \phi^2 +\phi^4 
 \right] dx.
 \label{A22}
\end{align} 
    The action $\int L dt$ is replaced with
\be
\label{A24}
S= \int_0^T  d   {\mathcal T_0}   \int d   {\mathcal T_1} \,  L   \left(    \phi,  \frac{\partial  \phi}{\partial {\mathcal T_0} }, 
   \frac{\partial  \phi}{\partial {\mathcal T_1}}     \right).
\ee

We choose the trial function 
in the form
   \be
\phi= A \,  \cos (\omega {\mathcal T_0} +\theta ) \, \mathrm{sech}  \left(   \frac{x}{b}   \right),
\label{A23}
\ee
where $A, b$ and $\theta$ are functions of the ``slow" time variable $\mathcal T_1$
while $\omega = 2\pi/T$. 
(Note that $\phi$ does not have to be assumed small.)
The interpretation of the width $b$ is the same as in the Ansatz \eqref{A4} 
while $A$ represents the maximum of the oscillon's amplitude rather than the amplitude itself.
Unlike the previous trial function \eqref{A4}, the variable  $A$ in \eqref{A23} 
is assumed to remain positive at all times. The phase correction $\theta$ is a new addition to the set of collective coordinates; 
its significance will be elucidated later  (section \ref{Modulation}). 
The choice   of the spatial part of the Ansatz will also be discussed below (section \ref{choice}).

Once the explicit dependence on $x$ and $\mathcal T_0$ has been integrated away, 
equations \eqref{A22} and \eqref{A24} give an effective action 
\[
S=  T \int d  {\mathcal T_1}  \, \mathcal L
\]
with 
\begin{align}
\mathcal L= 
 (D A)^2 b+\left( \frac13+ \frac{\pi^2}{36} \right) \frac{ (Db) ^2 A^2}{b} + A D A D b     \nonumber   \\ + (\omega + D \theta)^2  bA^2
 - \frac{A^2}{3b}     -4 b A^2 + \frac12 bA^4
\label{A26}
\end{align}
and  $D= \epsilon \frac{\partial}{\partial \mathcal T_1}$.  
Two Euler-Lagrange equations  are
\begin{align}
D^2 A+ 4 A      -(\omega+ D\theta)^2A - \left( \frac13+\frac{\pi^2}{36} \right)  \frac{(D b)^2}{b^2} A   
  \nonumber \\ 
=
(1+ \frac32  \sigma ) A^3    - \left( 2 \sigma +\frac13 \right) \frac{A}{b^2}     \label{A27} 
\end{align} 
and 
\[
D    \left[ (\omega+ D \theta) b A^2 \right]=0. 
\]
The last equation can be integrated to give
\be
\label{A290} 
(\omega+ D \theta) b A^2 = \ell,
\ee
where $\ell$ is a constant of integration. 
Eliminating the cyclic variable $\theta$ between \eqref{A27} and \eqref{A290} we arrive at
\begin{subequations}
\label{2}
\begin{align} 
D^2 A- \left( \frac13+\frac{\pi^2}{36} \right)  \frac{(D b)^2}{b^2} A   + 4 A      -   \frac{\ell^2}{b^2A^3} 
  \nonumber \\ 
=
(1+ \frac32  \sigma ) A^3    - \left( 2 \sigma +\frac13 \right) \frac{A}{b^2}.   \label{A270} 
\end{align}
The third Euler-Lagrange equation for the Lagrangian \eqref{A26}   does not involve $\theta$:
\be
 D^2 b + 2 \frac{DA}{A} D b = 4 \sigma  \left( \frac{1}{b^2}- \frac34 A^2 \right) b.
 \label{A28}  
 \ee
 \end{subequations}
Equations \eqref{2} constitute a four-dimensional conservative system with a single control parameter $\ell^2$.

\subsection{Slow dynamics and stationary points} 
\label{Stat_points}

The oscillon
corresponds to a fixed-point solution of the system \eqref{2}. There are two coexisting fixed points
for each $\ell^2$ in the interval $(0, \frac{64}{9})$. We denote their components by $(A_+, b_+)$ and $(A_-, b_-)$, respectively. Here
\be
A_{\pm}^2= \frac83 \pm \sqrt{ \frac{64}{9}- \ell^2}, \quad
b_{\pm}^2= \frac43 \frac{1}{A_{\pm}^2}.
\label{C10}
\ee

Turning to the stability of these, 
 we note that all derivatives in equations \eqref{2} carry a small factor $\epsilon$. Accordingly, 
  most of the time-dependent solutions of that
system  evolve on a short scale $\mathcal T_1 \sim \epsilon$. This is inconsistent with our original 
assumption that  $\partial \phi / \partial \mathcal T_1=O(1)$.
There is, however, a particular $\ell$-regime
where solutions change slowly and the system \eqref{2} is consistent.
Specifically, slowly evolving nonstationary solutions 
 can be explicitly constructed in the vicinity of the  value $\ell_c^2= \frac{64}{9}$;
see Appendix  \ref{App}. 
This value   proves to be a
saddle-centre bifurcation point
 separating a branch of 
  stable equilibria, namely $(A_-, b_-)$, 
   from an unstable branch, $(A_+, b_+)$.

Since the asymptotic construction  presented in the Appendix is limited to the neighbourhood of  the bifurcation value $\ell_c$, 
we do not have access to the oscillon perturbations outside that parameter region. 
Nevertheless, it is not difficult to realise that 
the two fixed points  maintain their stability properties over their entire domain of existence, $0 \leq \ell^2 < \ell_c^2$. 
 Indeed, the stability  may only change as $\ell$ passes through the value $\ell_0$ given by a root of $\det M=0$,
where $M$ is the linearisation matrix. 
(The evolution is slow and the system \eqref{2} is consistent in the vicinity of that point.)
There happens to be  only one such root and it is given exacty by $\ell_c$; see Appendix \ref{App}.

In order to compare the  variational results to conclusions of the direct numerical simulations of
equation \eqref{A1}, we return to the 
 oscillon Ansatz \eqref{A23}.  Switching from the  parametrisation by $\ell$ to the frequency parameter $\omega$, 
 two branches of fixed points \eqref{C10} can be characterised in a uniform way:
\be 
\label{A30}
A= \frac{2}{\sqrt 3} \sqrt{4-\omega^2},  \quad  b=\frac{1}{\sqrt{4-\omega^2}}. 
\ee
(The relations \eqref{A30} result by letting $\ell= \omega bA^2$ in  \eqref{C10}.)
  The frequencies $\omega_c \leq \omega <  2$ correspond to stable oscillons and those in the interval $ 0 \leq \omega < \omega_c$ 
 to unstable ones. Here 
 \be
 \omega_c= \sqrt 2.
 \label{omega}
 \ee
 The third collective coordinate  in \eqref{A23} --- the  phase correction $\theta$  --- can be assigned an arbitrary constant value.

Note that the expressions \eqref{A30} 
agree with the asymptotic result \eqref{A2} in the $A, b^{-1} \to 0$ limit.

\subsection{Numerical verification} 

We simulated the partial differential equation \eqref{A1} 
using a pseudospectral numerical scheme with $2^{13}$ Fourier modes.
The scheme imposes periodic boundary conditions $\phi(L)=\phi(-L)$ and
$\phi_x(L)=\phi_x(-L)$, where the interval should be chosen long enough to prevent any radiation re-entry. 
(Our $L$ was pegged to the estimated width of the oscillon, varying between  $L=20$ and
$L=100$.)

              \begin{figure}[t]
 \begin{center} 
                          \includegraphics*[width=\linewidth] {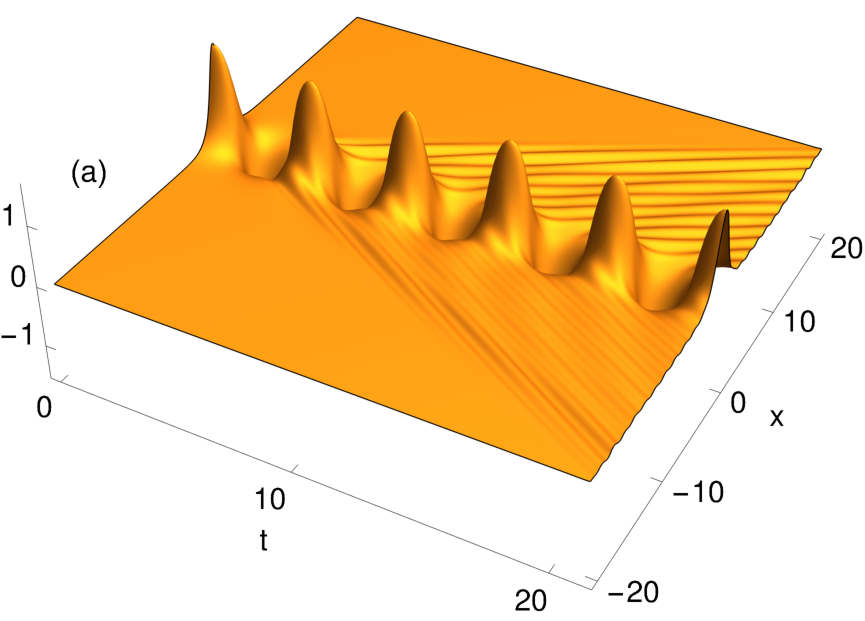}                          
                                     \includegraphics*[width=\linewidth] {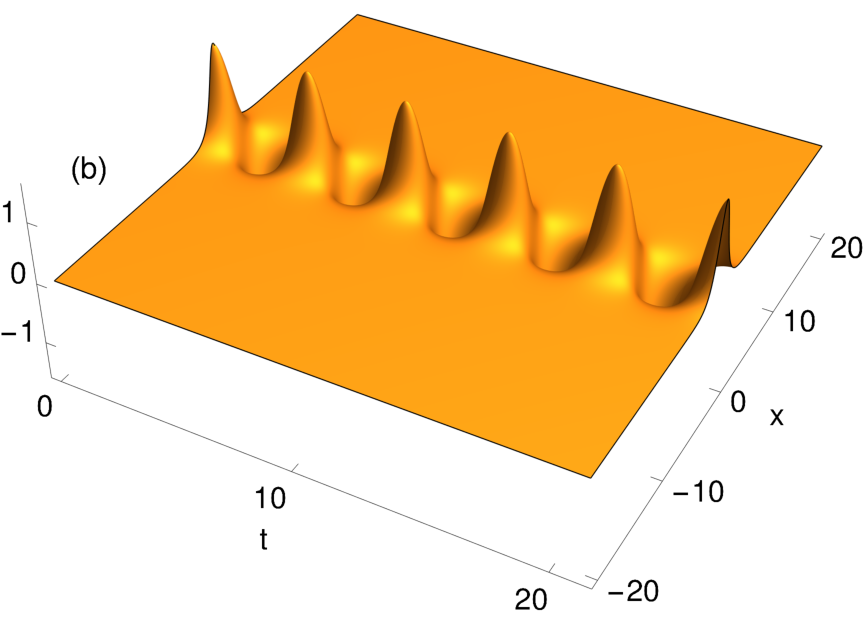}   
                                             \end{center}
  \caption{
  Top panel: 
    the Kosevich-Kovalev oscillon with
  $\omega =1.06  \,   \omega_c$
  (where $\omega_c=\sqrt 2$).   
  The oscillon is stable: despite the energy loss to radiation waves, 
 any changes in its period and  amplitude are hardly visible. 
   This figure is obtained by the numerical simulation of equation \eqref{A1}. 
   Bottom panel: the variational approximation \eqref{A23} with the matching $\omega$.
   Here $A$ and $b$ are as in \eqref{A30} with $\omega  = 1.06  \, \omega_c$, and
 $\theta=0$. Except for the absence of the radiation waves, the variational pattern is seen to be a good fit for the true oscillon.
     \label{3D}
  }
 \end{figure}

Using the initial data in the form 
\[
\phi(x,0)=  A_0 \sech \left(   \frac{x}{b_0}
\right),   
 \quad 
\phi_t(x,0)=0
\]
with $b_0= (2/ \sqrt 3)A_0^{-1}$ and
  varied $A_0$, we were able to create stable oscillons
with frequencies ranging from $\omega=1.03 \, \omega_c$ to  $\omega=2$. 
(Here  $\omega = 2 \pi/T$, where $T$ is the observed period of 
the localised periodic solution.) 
This  ``experimental" stability domain is in good agreement with the variational result $\omega_c \leq \omega< 2$.

The $3 \%$ discrepancy between two lower threshold values  can be attributed to the emission 
 of radiation and 
  the oscillon's core deformation  due to the third harmonic excitation.
(The  presence of the third harmonic in  the oscillon's core is manifest already in the asymptotic solution \eqref{A2}.)
The radiation intensifies and deformation becomes more significant as the oscillon's amplitude grows (Fig \ref{3D}(a));
yet the variational approximation disregards both effects (see Fig \ref{3D}(b)).

Once the evolution has settled to an oscillon with a  period $T$, we would measure its amplitude 
\be
A= \max_T  \left|\phi(x,t) \right|_{x=0}
\label{Q1}
\ee
and evaluate its width  which we define by
\be
b =  \frac{1}{2A^2} 
\max_T   
  \int_{-L}^L \phi^2(x,t)  dx.
  \label{Q2}
\ee
In \eqref{Q1}-\eqref{Q2},  the maximum is evaluated over the time interval $t_0 \leq t < t_0+T$, where 
 $t_0$ was typically chosen as the position of the third peak of $\phi(0,t)$.

\begin{figure}[t]
 \begin{center}  
                 \includegraphics*[width=\linewidth] {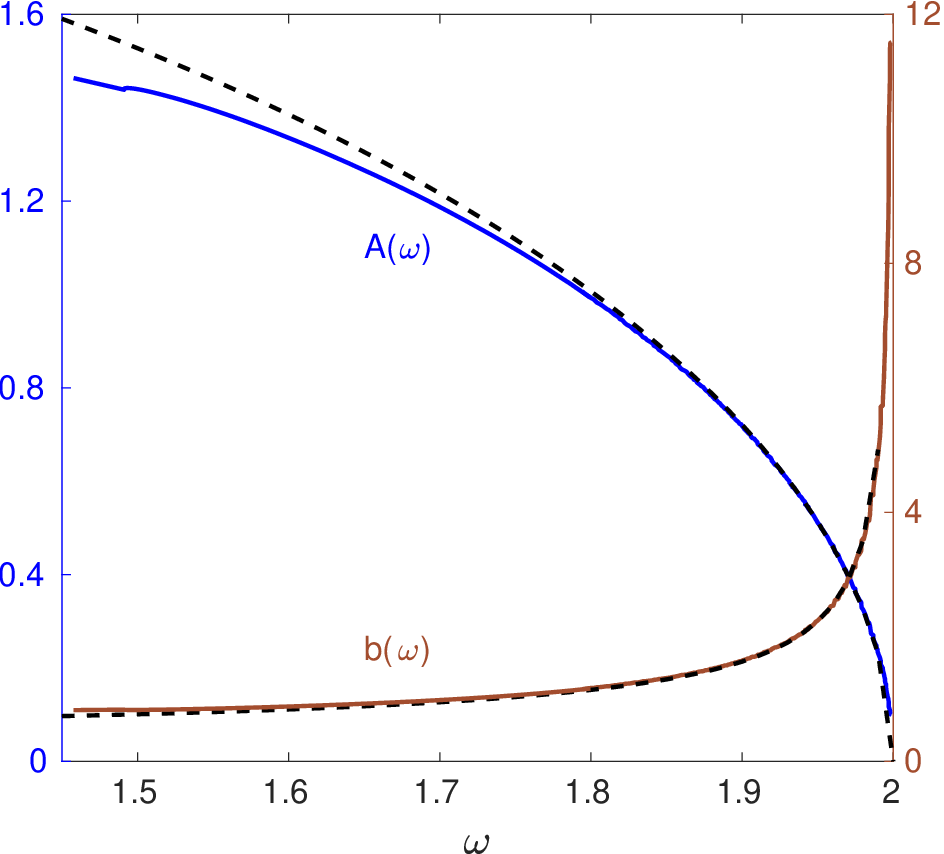}             
                              \end{center}
  \caption{The amplitude and width 
  of the oscillon as functions of its frequency. 
  The solid curves depict results of the numerical simulations of the partial differential equation \eqref{A1}.
  The blue curve traces the amplitude-frequency and the brown one gives the width-frequency dependence.
The nearby dashed lines describe the corresponding variational approximations \eqref{A30}.
      \label{A_omega_b_omega}
  }
 \end{figure}

Figure \ref{A_omega_b_omega} compares the amplitude and  width of the numerically generated oscillon 
with their variational approximations \eqref{A30}.
The difference between the numerical and variational results  grows as $\omega$  approaches $1.03 \, \omega_c$ 
--- yet the relative error in the amplitude remains below $8 \%$ and the  error in the width does not exceed 12.5\%.

    \section{Two remarks on the method}
    \label{Two_remarks}
    
\subsection{Modulation, instability and significance of $\theta$} 
\label{Modulation}

The inclusion of the 
 cyclic coordinate $\theta(T_1)$ is crucial for our variational  approach. 
To show that, we compare  the  system \eqref{2} incorporating, implicitly,  three degrees of freedom with
its two-degree ($A$ and $b$) counterpart.

Linearising equations \eqref{2} about the fixed point \eqref{A30}
and considering small perturbations with the time dependence $e^{(\lambda / \epsilon)  T_1} $, we obtain a characteristic equation 
\be
  \lambda^4+ (16-5A^2+ 3 \sigma A^2) \lambda^2 - 18 \sigma A^2 \left(A^2- \frac83 \right) =0.
 \label{P2}
\ee
When $A^2$ is away from 0 or $8/3$, all eigenvalues $\lambda$ are of order 1. This means that 
contrary to the  assumption under which the system \eqref{2} was derived, small perturbations evolve on a short scale $T_1 \sim \epsilon$ rather than $T_1 \sim 1$. 
The variational method cannot provide trustworthy information on the stability or modulation frequency of the oscillons 
with those $A$.

There are two regions where 
a pair of $O(\epsilon)$-eigenvalues  occurs and, consequently, our approach is consistent.
One region consists of small $A \sim \epsilon$; this range accounts for the asymptotic regime \eqref{A2}.
The second region is defined by $|A^2-8/3| =O(\epsilon^2)$ 
or, equivalently, by  $| \omega-\omega_c| \sim \epsilon^2$. 
As $\omega$ is reduced through $\omega_c$, a pair of opposite imaginary eigenvalues converges at the origin and moves onto the 
positive and negative real axis:
\[
\lambda^2= -\frac{16 \sqrt{2} \sigma}{\sigma + 1/3} (\omega-\omega_c) +O\left( (\omega-\omega_c)^2 \right). 
\]
At this point, a slow modulation of the principal harmonic $\cos (\omega_c t)$ with the modulation frequency $\sim (\omega-\omega_c)^{1/2}$ gives way to
an exponential growth of the perturbation. (For an  explicit construction of the time-dependent solutions of the system  \eqref{2}, see Appendix \ref{App}.)

Had we not included $\theta(T_1)$ in our trial function --- 
that is, had we set $\theta=0$ in equation \eqref{A23} --- we would have ended up with the same fixed point \eqref{A30} but
a different characteristic equation:
\be 
\lambda^4 + (3 \sigma-2) A^2 \lambda^2 -9 \sigma A^4=0.
\label{A32}
\ee
Equation \eqref{A32} does not  have roots of order $\epsilon$ outside the asymptotic domain $A \sim \epsilon$. 
Therefore, the multiscale variational Ansatz excluding the cyclic variable $\theta(T_1)$ is inconsistent with
the  slow evolution  of the collective coordinates $A(T_1)$ and $b(T_1)$.

\subsection{Insensitivity to  spatial shape variations}
\label{choice}

The $x$-part of the  trial function \eqref{A23} was chosen so as to reproduce the asymptotic representation
\eqref{A2} and match the  amplitude-frequency relationship as $\omega \to 2$. 
As for the global behaviour of the $A(\omega)$ curve, the variation of the spatial profile of the trial function  has little effect on it
--- as long as the function remains localised.

To exemplify  this insensitivity to the Ansatz variations, we replace the exponentially localised trial function \eqref{A23} with a gaussian:
\be
\label{E1}
\phi= A \,  \cos (\omega {\mathcal T_0} +\theta ) \, e^{-  \left( x/b  \right)^2}.
\ee
As in \eqref{A23}, 
the amplitude $A$,   width $b$ and  phase shift $\theta$ are assumed to be functions of the  slow time variable $\mathcal T_1=\epsilon  t$.   
Substituting in \eqref{A24} gives an effective action with the Lagrangian
\begin{align}
\mathcal L= 
 (D A)^2 b+\frac34 \frac{ (Db) ^2 A^2}{b} + A D A D b     \nonumber   \\ + (\omega + D \theta)^2  bA^2
 - \frac{A^2}{b}     -4 b A^2 + \frac{3 \sqrt 2}{8} bA^4.
\label{E2}
\end{align}
(Here, as before, $D = \epsilon \partial/ \partial T_1$). 
Equation \eqref{E2} has the same form as \eqref{A26} with the only difference residing in the value of some of the coefficients.

The Euler-Lagrange equations resulting from \eqref{E2} have a fixed-point solution
   \be 
\label{E3}
A=        \frac{2^{7/4} }{ 3} \sqrt{4-\omega^2},  \quad  b= \frac{1}{\sqrt 3} \frac{1}{\sqrt{4-\omega^2}}.
\ee
Note that the gaussian amplitude and width are related to $\omega$ by exactly same laws as the amplitude and width of the secant-shaped approximation
(equations \eqref{A30}). 
If    $A_g$ stands for
the amplitude \eqref{E3} and $A_s$ for the secant-based  result \eqref{A30}, 
the ratio $A_g(\omega)/ A_s(\omega) $ is given by  $\sqrt[4]{8/9} \approx 0.971$. 
Thus the   gaussian-based  amplitude-frequency curve reproduces the qualitative behaviour of  the curve \eqref{A30}, 
with the gaussian
 amplitude  being only 3\%-different from 
the amplitude of the secant-shaped variational oscillon.

Linearising the Euler-Lagrange equations about the fixed point \eqref{E3} we 
obtain a gaussian analog of the characteristic equation \eqref{P2}:
\be
  \lambda^4+ \left(   16-   \frac{27 \sqrt 2}{8}  A^2   \right) \lambda^2 - \frac{27}{8} A^2 \left(A^2- \frac{32}{9 \sqrt 2}  \right)=0.
 \label{E4}
\ee
The critical value of $A^2$ above which  a pair of opposite eigenvalues moves onto the real axis  is
 $32/9 \sqrt 2$. Remarkably, the corresponding threshold frequency $\omega_c=\sqrt 2$ coincides with the 
 value \eqref{omega} afforded by the secant Ansatz.

  \section{Conclusions}
\label{conclusions}

  This study was motivated by the numerous
 links and similarities between the Klein-Gordon oscillons
 and solitons of the  nonlinear Schr\"odinger equations. 
 A simple yet powerful  approach to the Schr\"odinger 
 solitons exploits the variation of action.
By contrast, the variational analysis of
the Klein-Gordon oscillons 
has not been  nearly as successful.

One obstacle to the straightforward (``naive") variational treatment of the oscillon is
that its width proves to be unsuitable as a collective coordinate in that approach.
The soliton's amplitude and width comprise  a standard choice
of variables in the Schr\"odinger domain, but making a similar  choice in the Klein-Gordon Lagrangian results in a singular four-dimensional system. 

This paper presents a variational method free from singularities. 
 The method aims at determining the oscillon's parameters, domain of existence and stability-instability transition points. 
The proposed formulation is based  on  a fast harmonic Ansatz supplemented by
 the adiabatic evolution of the oscillon's collective coordinates. 
 An essential component of the set of collective coordinates is the ``lazy phase":
a cyclic  variable accounting for nonuniform phase acquisitions. 



We employed the Kosevich-Kovalev model as a prototype equation exhibiting oscillon solutions. 
Our variational method establishes the oscillon's domain of existence ($0< \omega <2$) and identifies the 
frequency $\omega_c$ at which the oscillon loses its stability ($\omega_c = \sqrt 2$). 
The predicted stability domain is in good agreement with numerical simulations of the partial
differential equation \eqref{A1} which yield stable oscillons with frequencies $1.03  \,\omega_c \leq \omega <2$. 
The  variational amplitude-frequency and width-frequency curves  are consistent with
the  characteristics of the numerical solutions.

\section*{Acknowledgments}
Discussions with   Alexander Kovalev   are
 gratefully acknowledged. 
This study was supported by a collaboration grant from the National Research Foundation  of South Africa and Joint Institute for Nuclear Research
(NRF grant No.120467).

  \appendix 
  \section{Slow evolution near the onset of instability } 
  \label{App}

  The aim of this Appendix is to construct a slowly changing solution of the system \eqref{2} consistent with 
   the assumption used in the derivation of that system. The construction is carried out in the vicinity of 
  the parameter value signifying the onset of instability of the fixed point.

  We let 
\be
\label{C1}
\ell^2= \ell_0^2-  \epsilon^4,
\ee
where
$\ell_0$ is the parameter value to be determined.
The unknowns  are expanded as
 \begin{align}
 A= A_0+ \epsilon^2 A_1 + \epsilon^4 A_2 +...,    \nonumber \\
 b=b_0+\epsilon^2 b_1+ \epsilon^4 b_2+ ... .
 \label{C2} 
 \end{align}   
 Here $(A_0, b_0)$ is either of the two fixed points \eqref{C10}
  corresponding to $\ell=\ell_0$. Substituting \eqref{C1}-\eqref{C2} in \eqref{2} we equate coefficients of 
 like powers of $\epsilon$.

The order $\epsilon^2$ gives
\[
M {\vec Y}_1=0,
\]
where the matrix $M$ has  the form
 \be
\left( \begin{array}{cc}
4+ \frac{9 \ell_0^2}{4 A_0^2} -\frac{11 + 12 \sigma}{4}A_0^2   &    \left[ \frac32 \frac{\ell_0^2}{A_0^2} - \frac{1+6\sigma}{2}  A_0^2\right]           \frac{A_0}{b_0}              \\
6 \sigma A_0 b_0 & 6 \sigma A_0^2 
\end{array}
\right)
\label{M}
\ee
and the vector ${\vec Y}_1$ consists of the linearised perturbations of the fixed point:
\[
{\vec Y}_1=
\left( \begin{array}{c}
A_1 \\ b_1
\end{array}
\right).
\]

Setting $\det M=0$ determines the  value of $\ell_0^2$. This value turns out to coincide with $\ell_c^2$,
 the endpoint of the 
interval of existence of the fixed points:
\be
\ell_0^2=
\ell_c^2= \frac{64}{9}. 
\label{root}
\ee
As $\ell$ approaches  $\ell_c$,
the fixed points $(A_+, b_+)$ and $(A_-, b_-)$ join to become $(A_0,B_0)$.  Here
\be 
\label{C11}
A_0= \sqrt{\frac83}, \quad b_0= \frac{1}{\sqrt 2}.
\ee
 The components of the null eigenvector ${\vec Y}_1$ are readily identified: 
 \[
  A_1= A_0 y, \quad
 b_1= - b_0 y.
 \]
 Here $y=y(\mathcal T_1)$ is an arbitrary scalar function that will be determined at the next order of the expansion.

 At the order $\epsilon^4$ we obtain
 \be
 \label{C5}
  M {\vec Y}_2= {\vec F_2},
  \ee
   where
  \[
  {\vec F_2}= 
   \left( \begin{array}{c}
f_2 \\ g_2
\end{array}
\right)
\]
with
   \begin{align*}
   f_2 = -A_0  \partial_1^2 y+ 8 A_0\left(\frac43-\sigma\right) y^2 - \frac{1}{A_0^3 b_0^2},
   \\
   g_2 = b_0 \partial_1^2 y + 16 \sigma b_0 y^2.
   \end{align*} 
   The solvability condition for equation \eqref{C5} is
   \be
   {\vec Z} \cdot {\vec F_2}=0,
   \label{C6}
   \ee
   where 
   \[
   {\vec Z}= \left( \begin{array}{c}
A_0 b_0  \\ \frac43 ( 1-\frac{1}{3 \sigma} )
\end{array}
\right)   
\]
is the adjoint null eigenvector of the matrix $M$. Substituting for $A_0$ and $b_0$ from \eqref{C11}, 
equation \eqref{C6} yields
\be
\label{C7}
\left( 1 + \frac{1}{3 \sigma} \right) \partial_1^2 y = 16 y^2- \frac{9} {16}.
\ee

The amplitude equation \eqref{C7} has the form of the second Newton's law for a classical particle moving in the potential 
\[
U(y)= \frac{9}{16} y-\frac{16}{3}y^3.
\]
The potential has two equilibria: a minimum at $y_-= - \frac{3}{16}$ and a maximum  at $y_+= \frac{3}{16} $. 
These correspond to the two
 fixed points of the system
\eqref{2}:  the minimum pertains to
$(A_-, b_-)$ and the maximum to $(A_+, b_+)$.
Accordingly, the point $(A_-, b_-)$ is stable and $(A_+, b_+)$ unstable.

The stable fixed point is surrounded by a family of closed orbits.
The corresponding periodic solutions of equation \eqref{C7} are expressible in Jacobi functions:
\[
y(\mathcal T_1)  =-\frac{k^2+1}{3} \mu + k^2 \mu \, \mathrm{sn}^2 \left( \sqrt{
\frac{8 \sigma \mu}{1+ 3 \sigma} 
}
\mathcal T_1, k  \right),
\]
where 
\[
\mu= \frac{9}{16} \sqrt{
\frac{k^2+1}{k^6+1}
} .
\]
The elliptic modulus $k$, $0 \leq k \leq 1$,  serves as the parameter of the family.

\end{document}